\documentclass[preprint,showpacs,floats,nofootinbib,superscriptaddress,color,
  showkeys]{revtex4}
\usepackage[dvips]{graphicx}

\newcommand{\bea}{\begin{eqnarray}}
\newcommand{\eea}{\end{eqnarray}}
\newcommand{\be}{\begin{equation}}
\newcommand{\ee}{\end{equation}}
\newcommand{\ba}{\begin{eqnarray}}
\newcommand{\ea}{\end{eqnarray}}

\newcommand{\nk}{{\bf      k}}
\newcommand{\np}{{\bf      p}}
\newcommand{\nq}{{\bf      q}}
\newcommand{\nr}{{\bf      r}}

\begin{document}

\title{
General study of superscaling in quasielastic $(e,e')$ and $(\nu,\mu)$ reactions using
the relativistic impulse approximation}

\author{J.A. Caballero}
\affiliation{
Departamento de F\'{\i}sica At\'omica, Molecular y Nuclear,
Universidad de Sevilla, 41080 Sevilla, SPAIN}

\date{\today}


\begin{abstract}
  The phenomenon of superscaling for quasielastic lepton induced reactions 
  at energies of a few GeV is investigated within the framework of the relativistic 
  impulse approximation. A global analysis of quasielastic inclusive electron and 
  charged-current neutrino scattering reactions on nuclei is presented.
  Scaling and superscaling properties are shown to emerge from both types of
  processes. The crucial role played by final state interactions is evaluated by
  using different approaches. The asymmetric shape presented by the experimental 
  scaling function, with a long tail in the region of positive values of the scaling
  variable, is reproduced when the interaction in the final state between the knockout
  nucleon and the residual nucleus is described within the relativistic mean field approach.
  The impact of gauge ambiguities and off-shell effects in the scaling
  function is also analyzed.
\end{abstract}

\pacs{25.30.Pt; 13.15.+g; 24.10.Jv}
\keywords{Electron scattering reactions, charged-current neutrino
  reactions, scaling, superscaling, quasielastic peak, relativistic mean field}

\maketitle

\section{Introduction}

Scaling is a very general phenomenon~\cite{West74} occurring in various areas of
physics that deal with probes weakly interacting with many-body
systems in which a single constituent in the target system absorbs
the energy and momentum transfer. The validity of the concepts of
scaling~\cite{Day90} and superscaling~\cite{DS199} applied to inclusive
quasielastic (QE) electron scattering at intermediate to high
energies has been investigated at depth in~\cite{DS199,DS299,MDS02}.
From an exhaustive analysis of the $(e,e')$ world data one
concludes that the scaling behaviour is highly fulfilled~\cite{DS299}. One
may distinguish between scaling of the first kind,
which corresponds to reduced cross sections being independent on
the momentum transfer $q$, and scaling of second kind, namely, no 
dependence on the nuclear species. The analysis of data shows that
scaling of first kind is reasonably well respected at
excitation energies below the QE peak, usually called the scaling region, 
whereas scaling of second kind is excellent in the same region. 
The simultaneous occurrence of both kinds of
scaling is named superscaling.
At energies above the QE peak, where nucleon resonances are important, both
types of scaling, first and, to a lesser extent, second are broken. 
Scaling violations are shown to reside mostly in the transverse response, but
not in the longitudinal which appears to superscale~\cite{MDS02}. 
These results are in accordance
with the important contributions expected in the transverse channel due to
effects beyond the impulse approximation:
inelastic scattering~\cite{Alvarez-Ruso:2003gj,BCDM04},
correlations and meson exchange currents (MEC) in both the 1p-1h
and 2p-2h sectors~\cite{Amaro:2001xz,Amaro:2002mj,Amaro:2003yd,DePace:2003xu,DePace}.

The scaling analysis of QE $(e,e')$ data was extended
into the $\Delta$ region~\cite{neutrino1},
leading to the extraction of two different scaling functions which
embody the nuclear dynamics in the two regions. The scaling
approach has been exploited to predict inclusive charged-current (CC)
neutrino-nucleus cross sections. This strategy is based on the assumption
that a universal scaling function exists, which is
valid for both electron and neutrino scattering reactions,
provided that the corresponding kinematics are similar.
This hypothesis of a universal scaling function, which is true by construction in the relativistic
Fermi Gas (RFG) model~\cite{scaling88}, 
was further investigated when final state interactions (FSI)
were included. The analysis performed in~\cite{neutrino2} within the context of
a non-relativistic mean field calculation, but
incorporating important aspects of relativity at the level of the current operators,
called semirelativistic (SR) approach, proved that scaling properties were
highly fulfilled for the electromagnetic responses at intermediate to high energies. 
Moreover, the scaling function extracted from these coincides with that from the
$(\nu,\mu)$ reaction. However, it presents a symmetric shape which is not supported
by the analysis of $(e,e')$ data. This is not
unexpected because additional dynamical effects, which are beyond the non-relativistic
mean field picture considered in~\cite{neutrino2}, 
are needed in order to reproduce the {\it asymmetry} extracted from the experiment.

An investigation of the QE scaling properties of
CC neutrino-nucleus scattering within the
context of the relativistic impulse approximation (RIA) has been
presented in~\cite{PRL}. Although resorting only to one-body excitations, 
the RIA has been shown to provide the required asymmetry of the
scaling function when strong relativistic potentials are included in the model.
This makes an important difference with previous non-relativistic, or SR, calculations
based on the impulse approximation~\cite{neutrino2,Co,Ble01}.
The superscaling function evaluated from
QE $(\nu,\mu)$ calculations was compared with the $(e,e')$ phenomenological
one, showing the capability
of RIA models to yield the required properties of data.

In this paper we extend the investigation on scaling by performing
a global analysis of $(e,e')$ and $(\nu,\mu)$ reactions within the RIA framework.
We follow the general procedure of scaling and superscaling 
studies~\cite{MDS02,neutrino1,neutrino2,PRL}.
First, we calculate inclusive cross sections within
a specific model and then, obtain scaling
functions by dividing them by the relevant single-nucleon cross
sections weighted by the corresponding proton and
neutron numbers~\cite{MDS02,Barbaro:1998gu}. 
The scaling function so obtained is plotted against the 
scaling variable $\psi(q,\omega)$, and its scaling properties analyzed, i.e., 
we explore its dependence on the transfer momentum (first kind scaling) and
on the specific target nucleus (second kind scaling).

A detailed study of the off-shell effects and gauge ambiguities in the scaling function is
also presented. The analysis
performed and its comparison with data give us important clues on the validity of the
different theoretical descriptions considered. Furthermore, the consistency between $(e,e')$ and
$(\nu,\mu)$ calculations, which reflects the universality property of the superscaling 
function, is also clearly illustrated. Finally, we show that the RIA approach, in spite of its
simplicity, gives rise to the required shape presented by the {\it experimental} scaling 
function. Scaling and superscaling ideas have
been carried a step further to include neutral-current (NC) neutrino-nucleus scattering 
processes in~\cite{NC}. Here, NC differential cross sections were obtained by making use of the
{\it phenomenological} $(e,e')$ scaling function, i.e., the universality property was assumed.
It will be very interesting to investigate scaling for NC reactions
within the RIA framework. This will allow us to prove if various RIA-FSI models superscale,
and moreover, if the {\it universal asymmetric} scaling function emerges also from the
RIA calculations on NC processes.

The paper is organized as follows. In Section II we present a brief summary of the basic formalism
involved in describing inclusive QE electron-nucleus and CC neutrino-nucleus scattering processes.
We restrict ourselves to the plane wave Born approximation (PWBA) and assume the RIA.
The analysis of scaling and superscaling with the general expressions implied
is discussed in Section III. Here we also show the {\it experimental} scaling function together
with a phenomenological fit and a comparison with the simple RFG result. In Section IV we present
our discussion of the results. We start with a global analysis of QE $(e,e')$ reactions where
off-shell effects in the differential cross sections are investigated at depth. We continue with
the study of scaling properties showing that our theoretical results do scale even when
strong relativistic potentials are present. A separate analysis of the
different channel contributions is also presented. Comparison with data shows that our model
calculations do agree with experiment for specific descriptions of FSI.
To conclude, we compare the scaling functions evaluated from $(e,e')$ calculations
with those obtained from $(\nu,\mu)$ reactions~\cite{PRL}. Results show that they almost coincide, hence
the model is consistent with the fulfilment of the universality property. Finally, in Section V
we present our conclusions.

\section{Inclusive quasielastic lepton scattering formalism: the relativistic impulse approximation}

This work deals with lepton induced reactions at energies of a few GeV and QE kinematics.
In particular, we focus on inclusive electron scattering
and CC neutrino (antineutrino) scattering on nuclei and
assume the Born approximation (BA).
The leptonic variables (in the laboratory system) involved in the processes are 
$K^\mu=(\varepsilon,\nk)$ the 4-momentum of the incident lepton ($e$ or $\nu_\mu$)
beam, and $K'^\mu=(\varepsilon',\nk')$ the 4-momentum of the scattered lepton ($e'$ or $\mu$).
The process is mediated by the exchange of a virtual photon (electron
scattering) or a charged vector boson (CC neutrino scattering) with 4-momentum
$Q^\mu=(K-K')^\mu =(\omega,\nq)$. 

The general formalism for $(e,e')$ and $(\nu,\mu)$ reactions has been presented
in previous works~\cite{DS299,Amaro:2001xz,neutrino1,neutrino2,Alb97}.
Here we simply summarize those basic aspects needed
for later discussion. Assuming PWBA, i.e.,
one virtual particle exchanged and leptons described as free particles, the
QE differential cross section can be expressed in terms of separate nuclear response
functions. In the case of $(e,e')$ reactions we may write
\be
\left[\frac{d\sigma}{d\varepsilon'd\Omega'}\right]_{(e,e')}
=\sigma_M\left[v_LR^L(q,\omega)+v_TR^T(q,\omega)
\right]\, ,
\label{eq1}
\ee
where $\Omega'$ is the scattered electron solid angle and the term $\sigma_M$ represents
the Mott cross section. 
%
Analogously, for CC neutrino scattering reactions the differential cross section can be
written in the form~\cite{neutrino1,neutrino2}
\be
\left[\frac{d\sigma}{d\varepsilon' d\Omega'}\right]_\chi =
\sigma_0\left[ \widehat{v}_{CC} \widehat{R}^{CC}+
2\widehat{v}_{CL} \widehat{R}^{CL}+
\widehat{v}_{LL} \widehat{R}^{LL}+
\widehat{v}_{T} \widehat{R}^{T}+
2\chi\widehat{v}_{T'} \widehat{R}^{T'} \right]
\label{eq2}
\ee
with ($\varepsilon',\Omega'$) the muon kinematical variables. The symbol $\chi$
specifies neutrino-induced reactions ($\chi=+$) or antineutrino-induced
reactions ($\chi=-$) and the term $\sigma_0$ depends on the Fermi
constant and the Cabibbo angle (see~\cite{neutrino1} for its explicit expression).
%
The kinematical factors $v_K$ and $\widehat{v}_K$ 
come solely from the electromagnetic and weak leptonic tensors, respectively, and their
explicit expressions can be found in~\cite{DS299,Amaro:2001xz,neutrino1}.

The electromagnetic $R^K$ and weak $\widehat{R}^K$ response functions contain the
whole dependence on the nuclear vertex coupling and are expressed by taking the
appropriate components of the nuclear tensor~\cite{DS299,Amaro:2001xz,neutrino1}.
This involves the matrix elements of
the virtual photon or charged boson interaction with the nuclear electromagnetic or weak
current. 
%
The inclusive hadronic electromagnetic tensor reads
\begin{equation} \label{hadronic}
W^{\mu\nu}(q,\omega)= \overline{\sum_{i}}\int\!\!\!\!\!\!\!\sum_f\delta(E_f-E_i-\omega)
\langle f | \hat{J}^\mu_{em}(Q)|i\rangle^*
\langle f | \hat{J}^\nu_{em}(Q)|i\rangle\, ,
\end{equation}
where $|i\rangle$ describes the initial target state 
and $|f\rangle$ represents a specific many-body final nuclear state. The term 
$\hat{J}^\mu_{em}(Q)$ refers to the nuclear electromagnetic many-body current operator.
A similar expression to (\ref{hadronic}) should be written for the weak tensor 
$\widehat{W}^{\mu\nu}$ in terms
of the nuclear weak many-body current operator $\hat{J}^\mu_w(Q)$. The electromagnetic
tensor given in (\ref{hadronic}) is an exceedingly complicated object 
which includes all possible
final states that can be connected with the initial ground state through the action of the
many-body current operator. 

In this paper we restrict ourselves to the QE 
kinematic regime and we adopt the relativistic impulse approximation. 
Within the RIA the many-body nuclear current operator is simply given as a sum of
single-nucleon current operators that only couple the target ground state to scattering
states lying in the one-body knockout space. 
%
The RIA approach has been extensively applied in investigations of
exclusive electron scattering reactions~\cite{Udias1,Udias2,highp,Udias3}. 
Further details on the model for
neutrino-nucleus scattering reactions have been presented in~\cite{Alb97,Alb98,Chiara03,Cris06}.
Within the RIA framework the main ingredient needed to evaluate the electromagnetic and weak 
tensor is the single-nucleon current matrix element,
\be
\langle \hat{J}^\mu (Q)\rangle = \int d\nr e^{i\nq\cdot\nr}
\overline{\psi}_{F}(\np_F,\nr)\hat{\Gamma}^\mu \psi_B^{jm}(\nr)\, .
\label{sn}
\ee
Here $\psi_B^{jm}({\bf r})$ and $\psi_{F}(\np_F,\,{\bf r})$ are
the wave functions for the initial (bound) nucleon and
for the emitted nucleon, respectively, and $\hat{\Gamma}^\mu$ is the corresponding 
single-nucleon current operator for electron ($\hat{\Gamma}^\mu_{em}$) or weak CC neutrino
($\hat{\Gamma}^\mu_w$) scattering. 

We describe the bound nucleon states as self-consistent
Dirac-Hartree solutions, derived within a relativistic mean field (RMF) 
approach using a Lagrangian containing $\sigma$, $\omega$
and $\rho$ mesons~\cite{boundwf,Serot}. 
The outgoing nucleon state is described
as a relativistic scattering wave function. 
Different options have been considered.
First, the Relativistic Plane-Wave Impulse Approximation (RPWIA), namely, 
the description of the knockout nucleons by means of 
plane-wave spinors. Second, the effects due to FSI
between the ejected nucleon and the residual nucleus. In our model FSI
effects are described by
using Dirac equation solutions in presence of
relativistic potentials. This constitutes  the Relativistic
Distorted-Wave Impulse Approximation (RDWIA)~\cite{Udias1}.

The use of energy-dependent complex relativistic optical potentials fitted to
elastic proton scattering data has proven to be successful in
describing {\em exclusive} $(e,e'p)$ scattering reactions under
QE kinematics~\cite{Udias1,Udias2,highp,Udias3,eep_experiment}.
In this case of {\it exclusive} reactions, the optical potentials are built to reproduce the 
contribution from the elastic channel. 
For {\it inclusive} processes such as $(e,e')$ and $(\nu,\mu)$,
the contribution from the inelastic channels should be retained. 
Ignoring them would lead to an underestimation of the
inclusive cross section~\cite{Chiara03,Jin,Kim}. 
Multiple nucleon knockout effects have been treated in detail
within the context of the Green function method~\cite{Horikawa1980,Chinn89,Giusti2003,Meucci2004}.
A simple way of obtaining the inclusive strength 
within the RIA is to use purely
real potentials. We consider two choices for the real part. The first
uses the phenomenological relativistic optical potential from the
energy-dependent, $A$-independent parameterizations (EDAIC, EDAIO,
EDAICa) derived by Clark {\it et al.}~\cite{Clark}, but with their
imaginary parts set to zero. The second approach consists of describing the
outgoing nucleon by means of distorted
waves obtained with the same relativistic mean field used to
describe the initial bound nucleon states. We refer 
to these two FSI descriptions as real Relativistic Optical Potential (rROP) and
RMF, respectively. Dispersion relation and Green function 
techniques~\cite{Horikawa1980,Chinn89,Giusti2003,Meucci2004}
lead to results which are close
to those obtained in the impulse approximation 
with either the rROP~\cite{Giusti2003,Meucci2004} or the mean
field~\cite{Horikawa1980}. 


Concerning the current operator we make use of the relativistic free
nucleon expressions~\cite{neutrino1,Cris,Forest}. For electromagnetic $(e,e')$ processes the
three usual options, denoted as CC1, CC2 and CC3, are considered:
\ba
\left[\hat{\Gamma}^\mu_{\mathrm {CC1}}\right]^{p(n)}_{em} &=& (F_1^{p(n)}+F^{p(n)}_2)
\gamma^\mu-\frac{F_2^{p(n)}}{2m_N}(\overline{P}+P_N)^\mu
\label{eq1s4} \\
\left[\hat{\Gamma}^\mu_{\mathrm{CC2}}\right]_{em}^{p(n)} &=& F_1^{p(n)}\gamma^\mu + 
\frac{iF_2^{p(n)}}{2m_N}\sigma^{\mu\nu}Q_\nu 
\label{eq2s4} \\
\left[\hat{\Gamma}^\mu_{\mathrm{CC3}}\right]_{em}^{p(n)} &=& \frac{F_1^{p(n)}}{2m_N}\overline{P}^\mu + 
\frac{i(F_1^{p(n)}+F_2^{p(n)})}{2m_N}\sigma^{\mu\nu}Q_\nu \, ,
\label{eq3s4}
\ea
where $F_1^{p(n)}$ and $F_2^{p(n)}$ are the Pauli and Dirac proton (neutron)
form factors, respectively, that depend
only on $Q^2$, and the on-shell 4-momentum 
$\overline{P}^\mu =(\overline{E},\np)$ with $\overline{E}=\sqrt{p^2+m_N^2}$ 
and $\np$ the bound nucleon momentum
has been introduced. Note that the three operators are equivalent for free on-shell nucleons
(they are connected by the Gordon transformation). However, the RIA deals in general
with off-shell bound and ejected nucleons. Hence the three operators lead to different results.
Moreover, the current is not strictly conserved and uncertainties dealing with the election of
gauge also occur~\cite{Caballero-NPA,Udi-PRL,Cris04}. 

The relativistic charged weak current of the nucleon is given as 
$\hat{\Gamma}^\mu_w=\hat{\Gamma}^\mu_V-\hat{\Gamma}^\mu_A$, 
where the vector and axial-vector current operators read
\ba
\hat{\Gamma}^\mu_V&=&F_1^V\gamma^\mu+i\frac{F_2^V}{2m_N}\sigma^{\mu\nu}Q_\nu \label{vector}\\
\hat{\Gamma}^\mu_A&=&\left[G_A\gamma^\mu+\frac{G_P}{2m_N}Q^\mu\right]\gamma^5 \, \label{axial},
\ea
with $F_{1,2}^V$ the isovector nucleon form factors given in terms of the electromagnetic ones
as $F_{1,2}^V=F_{1,2}^p-F_{1,2}^n$. The axial-vector and pseudoscalar form factors are parameterized
as
\begin{eqnarray}
G_A &=& \frac{g_A}{1-Q^2/M_A^2}\label{ga}
\\
G_P &=& \frac{4m_N^2}{m_\pi^2-Q^2}G_A  \label{gp}
\end{eqnarray}
with $g_A=1.26$ and $M_A=1032$ MeV (see~\cite{ABM,Bernard}).

\section{Scaling and Superscaling at the quasielastic peak}

Detailed studies of scaling and superscaling for electron-nucleus cross section
have been presented in~\cite{DS199,DS299,MDS02}. The analysis of
the $(e,e')$ world data has shown the quality of the scaling
behavior: scaling of the first kind (no dependence
on the momentum transfer) is quite good at
excitation energies below the QE peak, whereas scaling of second
kind (no dependence on the nuclear species) works extremely well in the
same region. 
In this paper our aim is to investigate 
the QE scaling properties of electron-nucleus and
CC neutrino-nucleus scattering within the
context of the RIA. Assuming
various RIA models we prove that they do superscale, and we compare the
associated scaling functions with the $(e,e')$ phenomenological
one. 
%

In what follows we present the basic expressions needed to get the
scaling functions. Several choices have been proposed in the literature for
the appropriate scaling variable. Here, following the analysis of the
RFG model, we adopt the dimensionless variable denoted as $\psi'(q,\omega)$,
\begin{equation}
\psi^\prime\equiv\frac{1}{\sqrt{\xi_F}}\frac{\lambda^\prime-\tau^\prime}
 {\sqrt{(1+\lambda^\prime)\tau^\prime+
\kappa\sqrt{\tau^\prime(1+\tau^\prime)}}} \, ,
\label{eq11}
\end{equation}
where $\lambda^\prime\equiv (\omega-E_{shift})/2m_N$,
$\kappa\equiv q/2m_N$, $\tau^\prime\equiv \kappa^2-\lambda^{\prime
2}$, and $\xi_F\equiv\sqrt{1+(k_F/m_N)^2}-1$. 
The term $k_F$ is the Fermi momentum and the energy shift
$E_{shift}$, taken from~\cite{MDS02}, has been introduced to force the maximum 
of the cross section to occur for $\psi'=0$. As usual the notation $\psi$
refers to the scaling variable when $E_{shift}=0$.

For inclusive QE electron scattering processes, the superscaling function 
is evaluated by dividing the differential cross section (\ref{eq1}) by the
appropriate single-nucleon $eN$ elastic cross section weighted by the corresponding proton and
neutron numbers~\cite{DS299,MDS02,Barbaro:1998gu} involved in the process.
We may write
\be
f(\psi',q)\equiv k_F\frac{\left[\displaystyle\frac{d\sigma}{d\varepsilon'd\Omega'}\right]_{(e,e')}}
{\sigma_M\left[V_LG_L(q,\omega)+V_TG_T(q,\omega)\right]} \, .
\label{fscaling}
\ee
The scaling behaviour can be also analyzed by taking into account the separate 
electromagnetic longitudinal ($L$) and transverse ($T$) contributions. 
Thus the following scaling functions are introduced:
\ba
f_L(\psi',q)&\equiv& k_F\frac{R^L(q,\omega)}{G_L(q,\omega)} \label{flscaling}\\
f_T(\psi',q)&\equiv& k_F\frac{R^T(q,\omega)}{G_T(q,\omega)}\label{ftscaling} \, .
\ea
The single-nucleon functions $G_L$ and $G_T$ are given by
\ba
G_L &=& \frac{(\kappa^2/\tau)\left[\tilde{G}^2_E+\tilde{W}_2\Delta\right]}
{2\kappa\left[1+\xi_F(1+\psi^2)/2\right]} \, , \label{gl}\\
G_T &=& \frac{2\tau\tilde{G}^2_M+\tilde{W}_2\Delta}
{2\kappa\left[1+\xi_F(1+\psi^2)/2\right]} \label{gt}\, ,
\ea
where the function $\Delta$ reads
\be
\Delta=\xi_F(1-\psi^2)\left[\frac{\sqrt{\tau(1+\tau)}}{\kappa}+\frac{1}{3}\xi_F(1-\psi^2)
\frac{\tau}{\kappa^2}\right] \, .  \label{delta}
\ee
As usual one has
\be
\tilde{G}_E^2\equiv ZG_{Ep}^2+NG_{En}^2\, , \,\,\,\,\, 
\tilde{G}_M^2\equiv ZG_{Mp}^2+NG_{Mn}^2\, \,\,\,\,\,
\tilde{W}_2=\frac{1}{1+\tau}\left[\tilde{G}_E^2+\tau\tilde{G}_M^2\right] \, ,
\label{gsquare}
\ee
involving the proton and neutron form factors weighted by the proton and neutron numbers $Z$ and
$N$, respectively. 


\begin{figure}
\begin{center}
\includegraphics[scale=0.9]{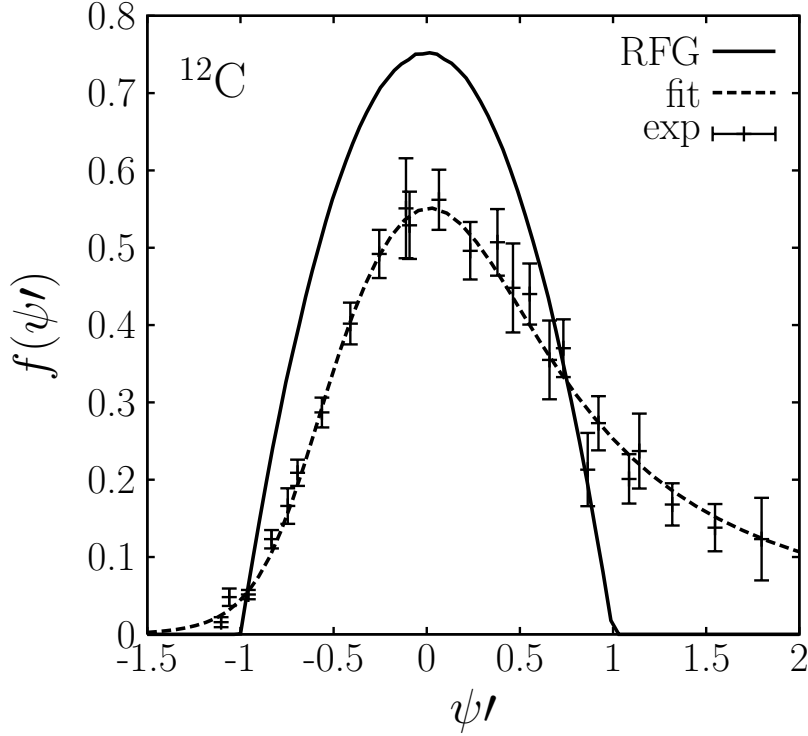}
\caption{RFG superscaling function compared to data and a parameterization of the results.}
\label{exp_rfg}
\end{center}
\end{figure}

At sufficiently high energies the function $f$ depends only on the scaling variable
$\psi'$ but not on the transferred momentum $q$. Moreover, $f(\psi')$ becomes also independent on
the momentum scale in the problem, that is, independent of $k_F$. The scaling behavior has
been clearly demonstrated from the analysis of the QE $(e,e')$ 
world data~\cite{DS199,DS299}.
The investigation of the separate contribution of the longitudinal and transverse
response functions has shown that scaling violations occur mainly in the $T$-channel due to
significant effects introduced by MEC, correlations and inelastic
scattering. From the global analysis presented in \cite{DS199,DS299} a universal {\it experimental} 
superscaling function $f_{exp}(\psi')$ has been defined. In Fig.~\ref{exp_rfg} we
present $f_{exp}(\psi')$ averaged over the
nuclei employed in the analysis, together with the corresponding fit. As noted, the 
{\it experimental} scaling function presents an asymmetric shape with a tail that extends
towards positive values of the scaling variable $\psi'$. This is in contrast with the RFG
superscaling function given by $f_{RFG}(\psi')=(3/4)(1-\psi'^2)\theta(1-\psi'^2)$, that
is symmetric, limited strictly to the region $|\psi'|\leq 1$ and with a maximum value of
$3/4$.

The general analysis of superscaling for CC neutrino-nucleus scattering has been presented in
previous works~\cite{neutrino1,neutrino2,PRL}. Here the superscaling function is obtained by dividing
the differential cross section evaluated within the RIA (\ref{eq2}) by
the corresponding weak single-nucleon cross section as given 
explicitly in Eqs.~(15,45,52,86-94) of \cite{neutrino1}. 
Details and specific expressions are also given in appendix C of \cite{neutrino2}.
This {\it theoretical} scaling function calculated from CC neutrino-nucleus cross sections can
be compared directly with the one corresponding to $(e,e')$ scattering calculations as well
as with $f_{exp}(\psi')$ shown above. 
This allows a check on the 
universality assumption of $f(\psi')$ and on the capabilities of different RIA models to yield
or not the required properties of the experimental scaling function.
By analogy to $(e,e')$, and
in addition to the scaling function obtained from the CC neutrino-nucleus cross section,
one may also construct the separate contributions given by the longitudinal $L$, transverse $T$ and
axial-vector transverse $T'$ responses, $f_{L,T,T'}(\psi')$. Here the $L$ weak response
includes the contribution from the three terms in (\ref{eq2}),
namely, $\hat{v}_{CC}\hat{R}^{CC}+2\hat{v}_{CL}\hat{R}^{CL}+\hat{v}_{LL}\hat{R}^{LL}$.
In next section we present a detailed study of scaling and superscaling properties for
$(e,e')$ and $(\nu,\mu)$ reactions within the scheme of the RIA.

\section{RESULTS}

In this work we consider the PWBA, i.e., 
Coulomb distortion
of the leptons is neglected. Checks made for light-to-medium nuclei
within the {\it effective momentum
approximation}~\cite{Alb97,Chiara03} show that these effects are within
a few percent for the high energy lepton kinematics considered in 
this work. In this sense, our general conclusions about scaling are not modified by them.
Obviously, caution should be drawn on the analysis of heavier nuclear systems 
where a careful description of Coulomb distortion effects in the leptons (electrons
and muons) involved in both processes $(e,e')$ and $(\nu,\mu)$ is required.


\subsection{Inclusive QE $(e,e')$ reactions}

\subsubsection{Differential cross sections}


\begin{figure}[htb]
\begin{center}
\includegraphics[scale=0.9]{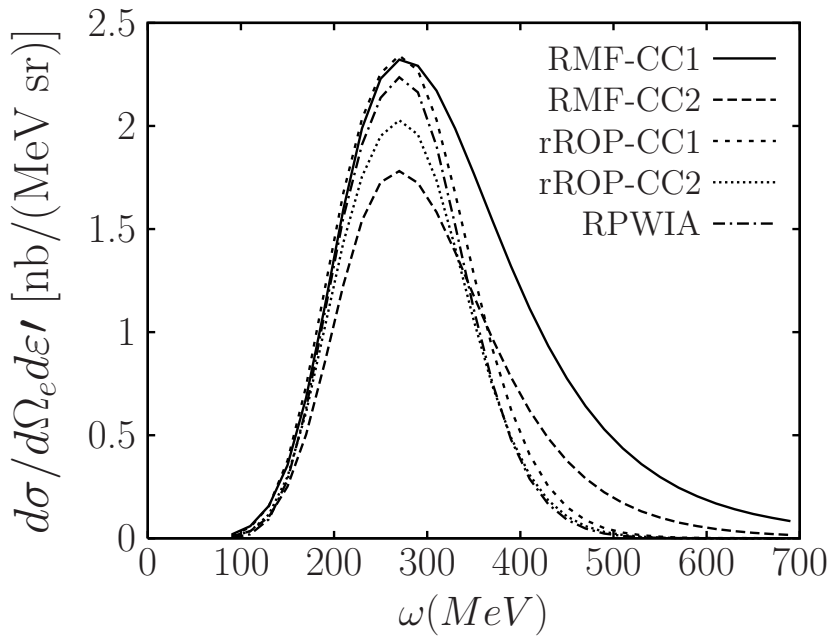}
\caption{Differential cross section for QE $(e,e')$ on $^{12}$C. FSI are described within
the RMF and the rROP models. Results correspond to CC1 and CC2 current operators. Also for reference
the RPWIA result is presented. The incident electron energy is $\varepsilon=1$ GeV and the
scattering angle $\theta_e=45^0$. See labels for reference on the different curves.}
\label{ria_cs}
\end{center}
\end{figure}

In what follows we present predictions for QE $(e,e')$ reactions on $^{12}$C within the
framework of the RIA. Results correspond to fixed values of the incident electron beam energy,
$\varepsilon=1$ GeV and scattering angle, $\theta_e=45^0$. The Lorentz gauge 
has been selected.
In next section we present results for the scaling function
concerning gauge effects.
In Fig.~\ref{ria_cs} we show the
differential cross section evaluated for the two current operators, CC1 and CC2, and FSI
included through the rROP and RMF potentials. As observed, the CC1 choice leads to a significantly
larger cross section, particularly within the RMF approach. Concerning FSI, the use of the
RMF potential gives rise to a clear asymmetry in the cross section with a pronounced tail extending
towards higher values of the transfer energy $\omega$. For reference we also include in
Fig.~\ref{ria_cs} the curve corresponding to RPWIA, i.e., no FSI. In such a case, the
effects introduced by the current operator choice are minor. The difference observed between
both FSI descriptions at high $\omega$-values is linked to the behaviour of the two 
relativistic potentials: 
whereas the RMF contains strong energy-independent scalar and vector potentials, 
the energy dependence
of the rROP makes its scalar and vector terms to be importantly reduced for high
nucleon kinetic energies (high transfer energy). Hence, rROP results get close to the
RPWIA ones for large $\omega$-values. 

The asymmetry in the differential cross section is proved to be
an effect entirely linked to the FSI description.
Only in presence of relativistic optical potentials with 
strong scalar and vector terms (RMF approach), 
a significant shift of strength to higher values of $\omega$ is 
shown to occur.
Details on the specific
mechanism that produces the asymmetric tail in the cross section 
within the RMF-FSI approach are given in~\cite{Udias-asymmetry}.


\begin{figure}[htb]
\begin{center}
\includegraphics[scale=0.8]{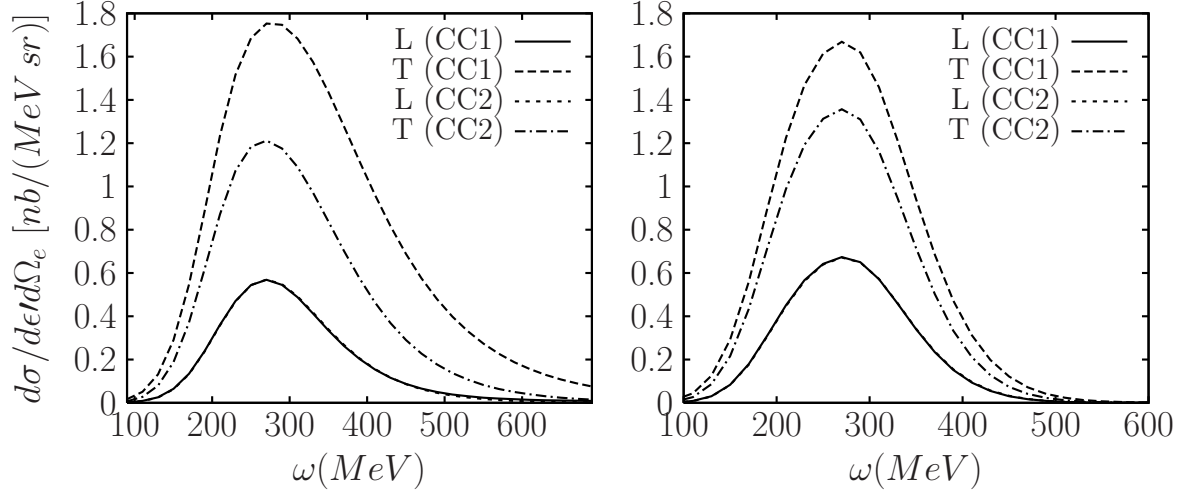}
\caption{Longitudinal ($L$) and transverse ($T$) contribution to the cross section. Left panel
corresponds to RMF description of FSI and right panel to rROP. In both cases results are presented
for CC1 and CC2 prescriptions.}
\label{RMF_ROP_LT}
\end{center}
\end{figure}

The importance of the current operator choice for the two FSI models is further
investigated in Fig.~\ref{RMF_ROP_LT}, where a separate analysis of the longitudinal 
and transverse contributions to the cross section is presented. 
As observed, the pure longitudinal response
is almost identical with both current operators. In the case in which FSI are neglected,
this has been also found previously~\cite{Caballero-NPA} and it was proven as due
to the validity of the Gordon transformation for the longitudinal contributions. 
For the case of the transverse
channel, the CC1 contribution is much larger than the CC2 (likewise for CC3) one.
Notice that the magnitude of this discrepancy depends on the specific FSI description, being
larger for RMF, whereas in the plane wave limit (RPWIA) both currents lead to very similar results.
The ambiguity introduced by the current choice for inclusive $(e,e')$ reactions has been already
signaled in some previous works~\cite{Horowitz,Chinn94}.


\begin{figure}[htb]
\begin{center}
\includegraphics[scale=0.7]{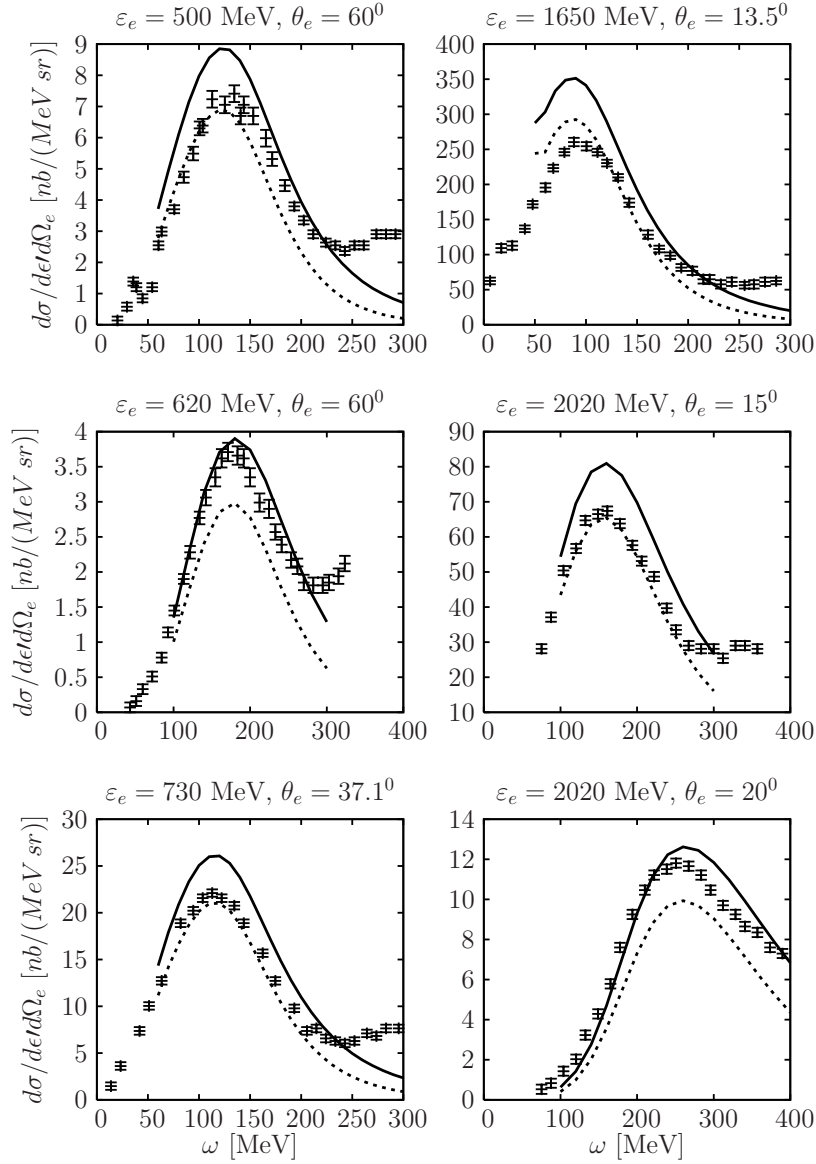}
\caption{Differential cross section for $(e,e')$ reactions on $^{12}$C. Theoretical predictions
correspond to the RMF-FSI approach with CC1 (solid line) and CC2 (dashed line). Different kinematics
(see labels) have been considered.}
\label{Exp_cs_C12}
\end{center}
\end{figure}

The large effects introduced by the current operator within the RMF (likewise for rROP) approach
can be also analyzed by comparing directly theoretical and experimental cross sections.
This may allow us to 
determine which particular choice is more appropriate. In Fig.~\ref{Exp_cs_C12}
we compare data~\cite{Butkevich,data1,data2,data3,data4,data5} with the results 
corresponding to the RIA-RMF approach with CC1 and CC2
at very different kinematics.
As a general rule we conclude that CC1 tends to overpredict data whereas the reverse 
applies to CC2. This outcome favours clearly the CC2 option as far as effects beyond the QE
peak ($\Delta$ and MEC effects) may also play a significant role 
in the analysis of the data even at the maximum of the peak. 
However, data are not conclusive yet as far as
the $\Delta$ and MEC contributions have not been evaluated.

\subsubsection{Analysis of scaling behavior}


\begin{figure}[htb]
\begin{center}
\includegraphics[scale=0.6]{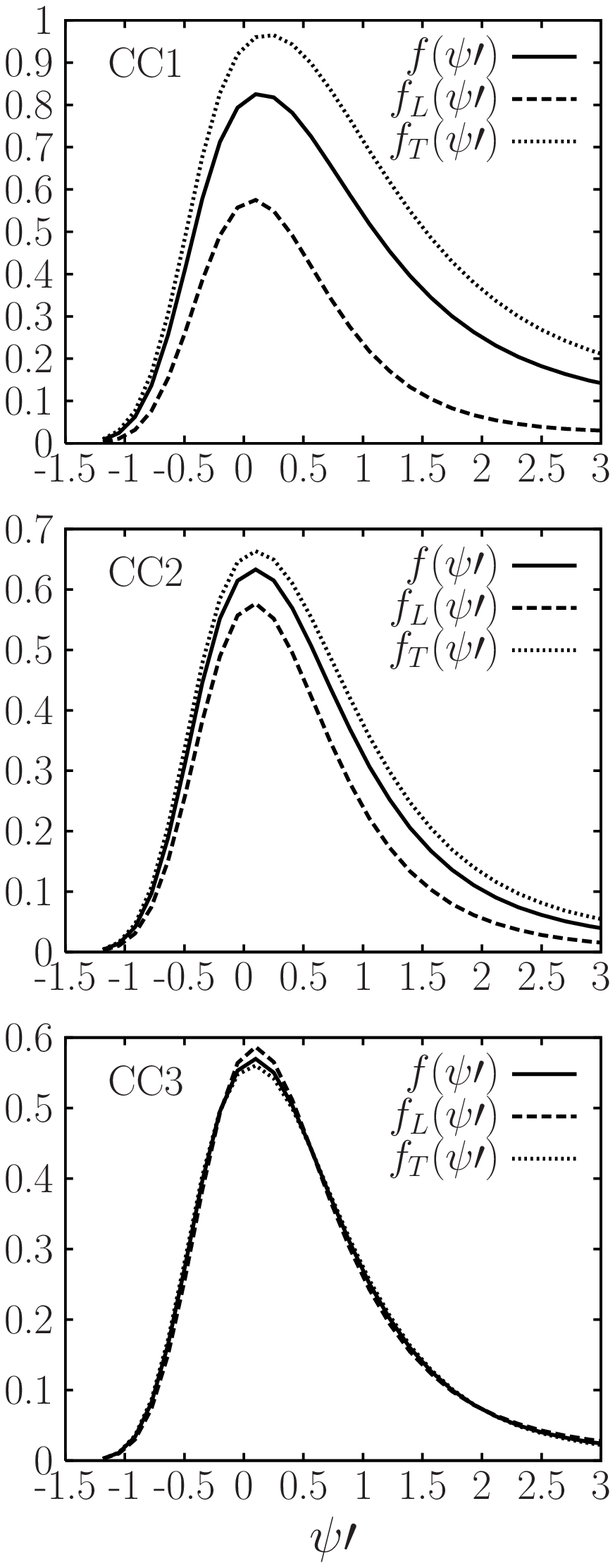}
\caption{Analysis of zeroth kind scaling. The global scaling function $f(\psi')$ (solid line) is compared
with the separate $L$ (dashed) and $T$ (dotted) contributions. All results correspond to RMF 
description of FSI and current operators: CC1 (top panel), CC2 (middle) and CC3 (bottom).}
\label{fLT_scaling}
\end{center}
\end{figure}

The important effects already shown in the differential cross sections are also visible
in the scaling function. In the following we rely on the investigation of the superscaling 
properties and present results
for the scaling function $f(\psi')$ (\ref{fscaling}), as well as the separate
longitudinal and transverse contributions, i.e., $f_L(\psi')$ and $f_T(\psi')$.
A comparison between the three scaling functions is presented in Fig.~\ref{fLT_scaling} where
for simplicity only the RMF-FSI model has been considered. Similar conclusions are drawn for the
rROP approach. In addition to the usual CC1 (top panel) and CC2 (middle) prescriptions we also
include the results for the CC3 choice (bottom panel). As observed, the difference between
$f(\psi')$ and the contributions $f_L(\psi')$ and $f_T(\psi')$ is very large for CC1, 
being much smaller for
CC2 and almost negligible for CC3. This means that zeroth kind scaling is fully broken 
for CC1 and RMF,
whereas only a mild (negligible) violation is observed for CC2 (CC3). 
This behaviour of the superscaling
function is in accordance with the very diverse contributions given in the $T$ channel by the
different current operators. In particular, the CC1 current and RMF description of FSI lead to
an important increase of the $T$ channel strength compared with the single-nucleon
contribution. On the contrary, the longitudinal function $f_L$ is found to be basically
the same for the three choices of the operator (likewise for rROP).


\begin{figure}[htb]
\begin{center}
\includegraphics[scale=0.7]{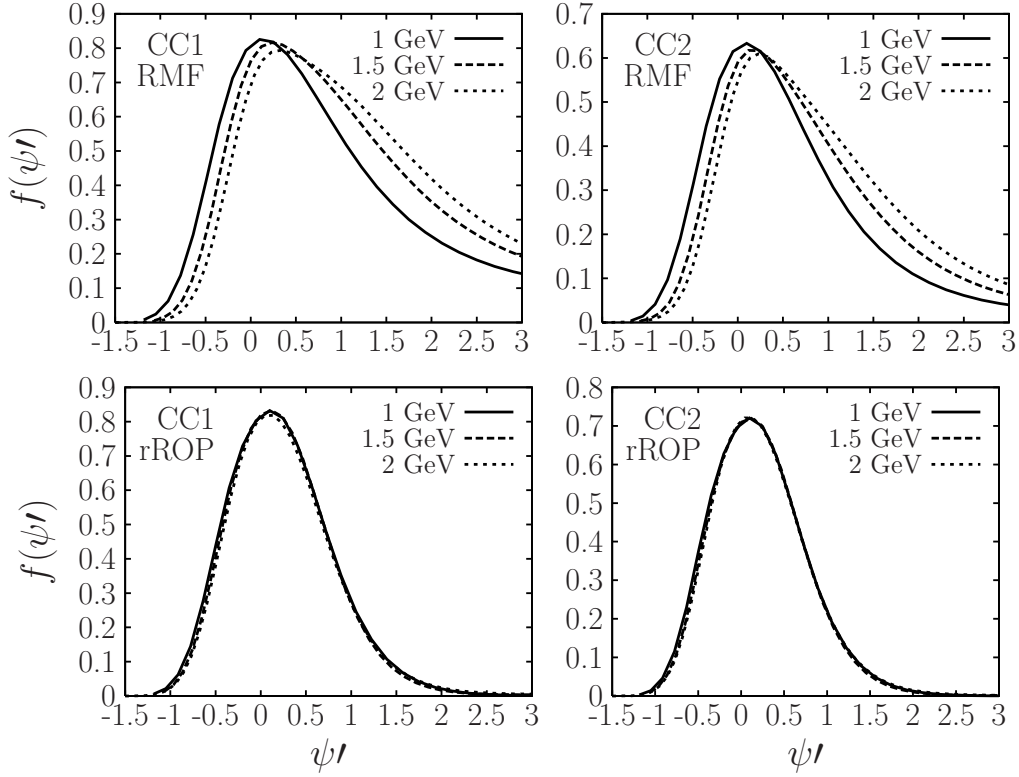}
\caption{Analysis of first kind scaling. Scaling function for three values of the incident
electron energy. Kinematics and target as in previous figures. Results correspond to RMF model
(top panels) and rROP (bottom) and the two choices of current operators: CC1 (left) and CC2 (right).}
\label{f_scaling_first}
\end{center}
\end{figure}

\begin{figure}[htb]
\begin{center}
\includegraphics[scale=0.7]{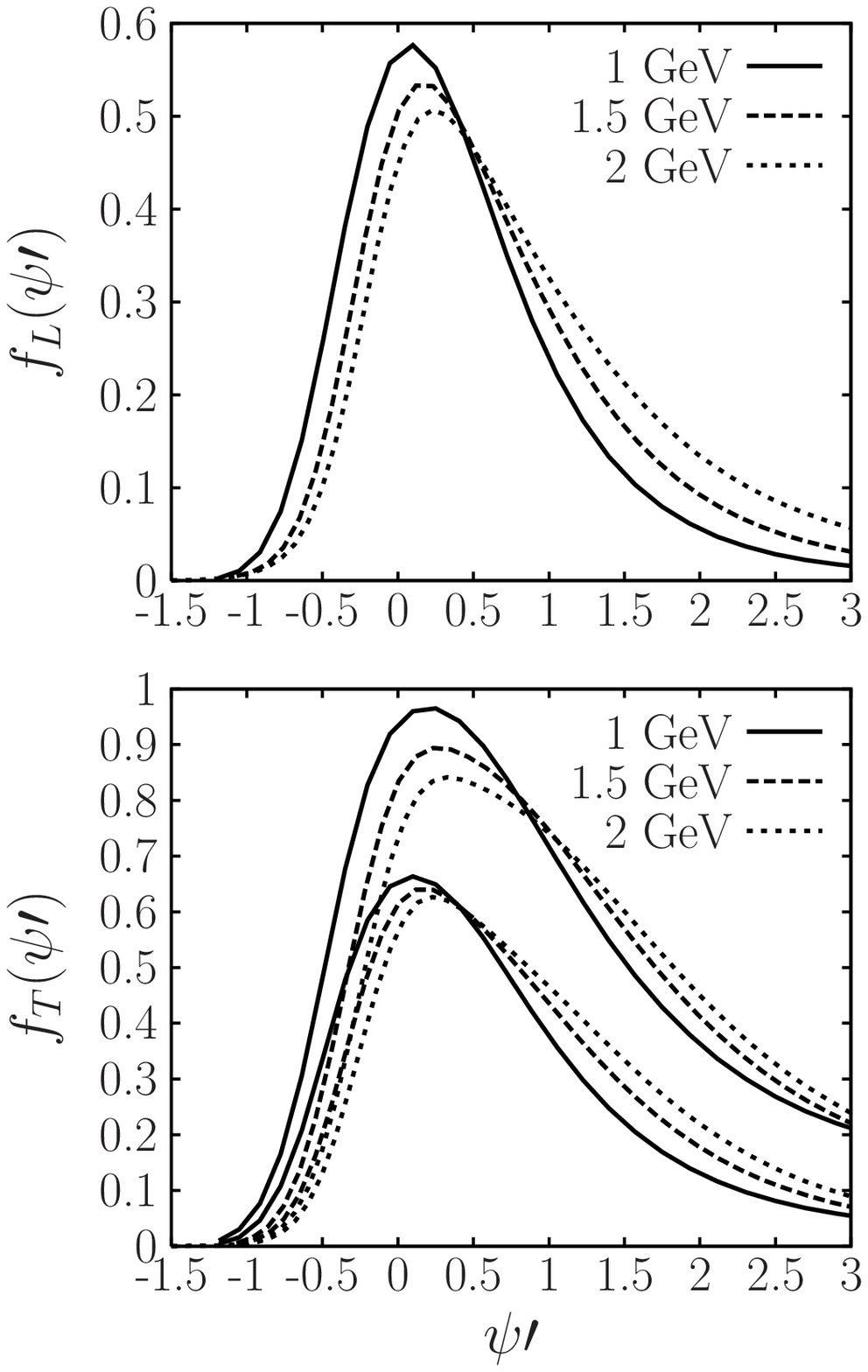}
\caption{As in previous figure but for the separate $L$ (top panel) and $T$ (bottom) contributions.
Only the RMF approach is considered. In the case of the $T$ response, the larger functions correspond
to the CC1 operator and the smaller ones to CC2. In the $L$ channel there is almost no distinction
between both operators.}
\label{f_scaling_LT_first}
\end{center}
\end{figure}

Scaling of first kind is explored in Fig.~\ref{f_scaling_first} where we present
$f(\psi')$ for three different values of the incident electron energy, $\varepsilon=1$,
$1.5$ and $2$ GeV. Results are shown for the two different descriptions of the FSI: 
RMF (top panels) and
rROP (bottom panels). In each case we make predictions for the two usual current prescriptions,
CC1 (left) and CC2 (right). As observed, the scaling function for the rROP model shows a very mild
dependence on the transfer momentum in both positive and negative $\psi'$ regions, i.e., scaling
or first kind is well satisfied. In the case of the RMF model, a slight shift occurs in the
``scaling region'' $\psi'<0$, whereas for $\psi'$-positive the model breaks scaling at roughly 
$30\%$. This violation is not in conflict with $(e,e')$ data that indeed leave room for some violation
of the first-kind scaling in this region, due partly to $\Delta$ production and partly to other
contributions, such as MEC and correlations.
A separate analysis of the $L$ and $T$ channels in the scaling function 
(see Fig.~\ref{f_scaling_LT_first}) shows that the breakdown of scaling within the RMF
description of FSI is similar for both channels independently of which current choice
is considered. In the case of rROP (and RPWIA) calculations scaling of first kind is excellent
for all current operators and in both channels.


\begin{figure}[htb]
\begin{center}
\includegraphics[scale=0.7]{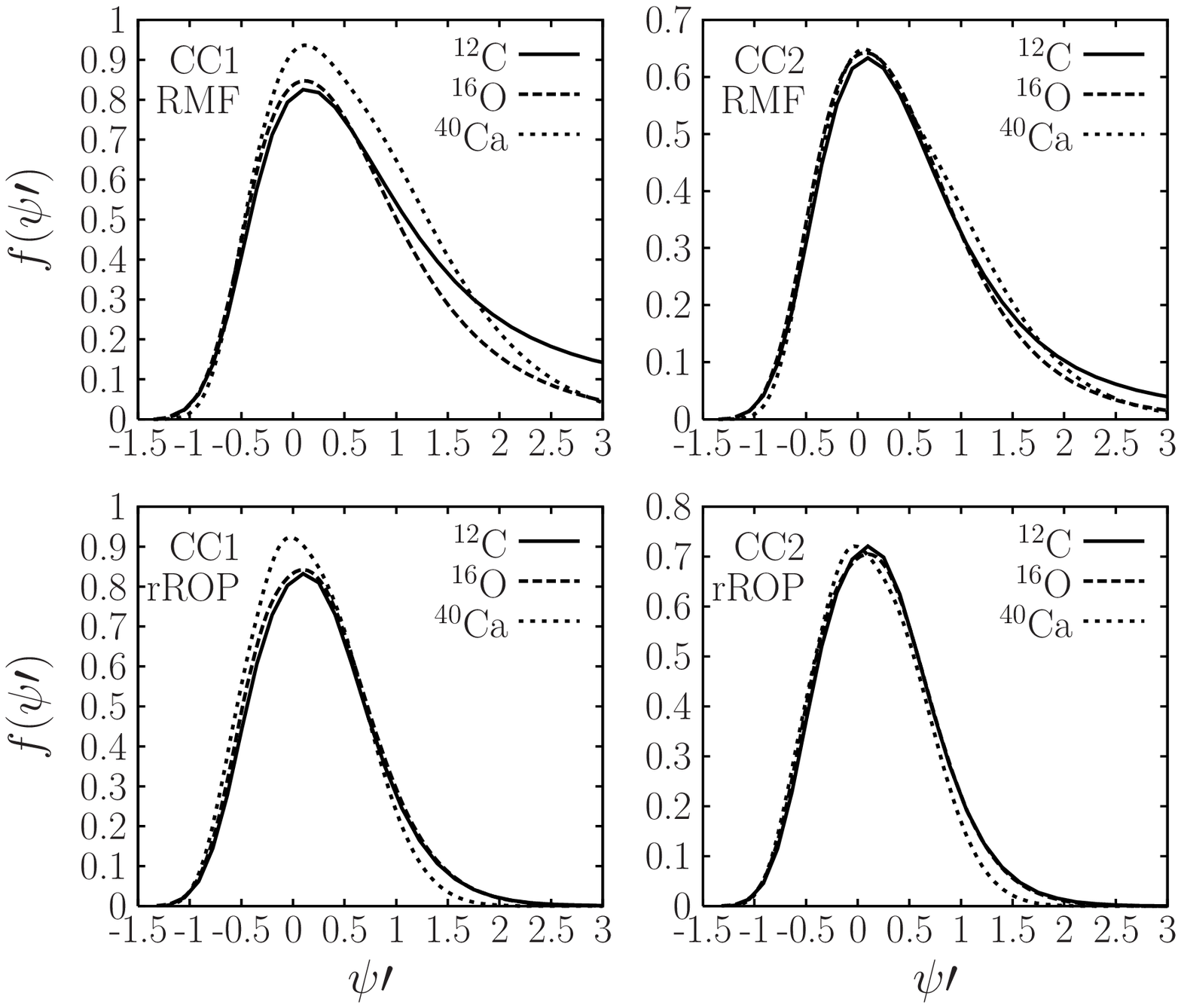}
\caption{Analysis of second kind scaling. Scaling function for three nuclei: $^{12}$C,
$^{16}$O and $^{40}$Ca. Kinematics as in previous figures. Results for the RMF (top panels) 
and rROP (bottom) models are presented for CC1 (left) and CC2 (right) prescriptions.}
\label{f_scaling_second}
\end{center}
\end{figure}

\begin{figure}[htb]
\begin{center}
\includegraphics[scale=0.7]{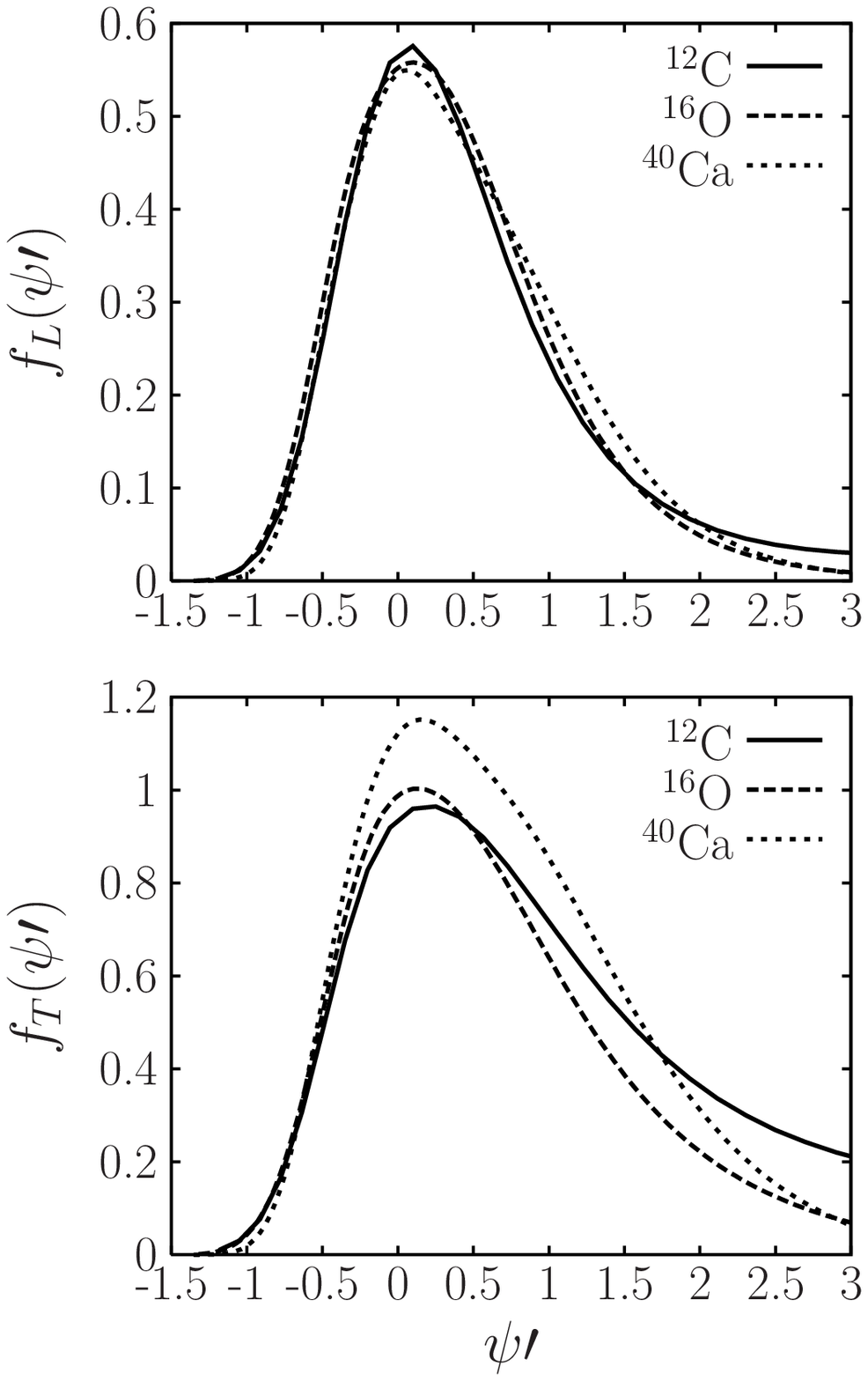}
\caption{Separate $L$ (top) and $T$ (bottom) contributions to second-kind scaling analysis. Results
correspond only to the RMF model.}
\label{f_scaling_LT_second}
\end{center}
\end{figure}

Scaling of second kind, i.e., independence on the specific nuclear system, is analyzed in
Fig.~\ref{f_scaling_second}. Here we present the results corresponding to CC1 (left panels)
and CC2 (right panels) operators and the two descriptions of FSI: 
RMF (top panels) and rROP (bottom panels).
In each case we compare the superscaling function evaluated for three different nuclei: $^{12}$C
(solid line), $^{16}$O (dashed) and $^{40}$Ca (short-dashed). The values of the Fermi momentum 
considered~\cite{MDS02} correspond to 
$k_F=216$ MeV/c ($^{16}$O), $k_F=228$ MeV/c ($^{12}$C) and $k_F=241$ MeV/c
($^{40}$Ca). As shown, the effects introduced by changing nucleus are very small for the
CC2 current choice and the two FSI descriptions. This is in complete accordance with data which
show that second-kind scaling is excellent. On the contrary, the CC1 operator leads to a significant
breakdown of scaling behaviour which affects both FSI descriptions. This violation of the
second kind scaling comes totally from the transverse response. This is clearly illustrated in
Fig.~\ref{f_scaling_LT_second} where we present results for the separate functions
$f_L(\psi')$ (top panel) and $f_T(\psi')$ (bottom) corresponding to the RMF model. Similar results
are obtained for the rROP approach. From these theoretical results and 
the exhaustive analysis of the $(e,e')$ world data, which proves
the excellent quality of second-kind scaling behaviour,
one may question the validity of the CC1 operator used in describing QE $(e,e')$ processes.

\subsubsection{Comparison with experiment}


\begin{figure}[htb]
\begin{center}
\includegraphics[scale=0.9]{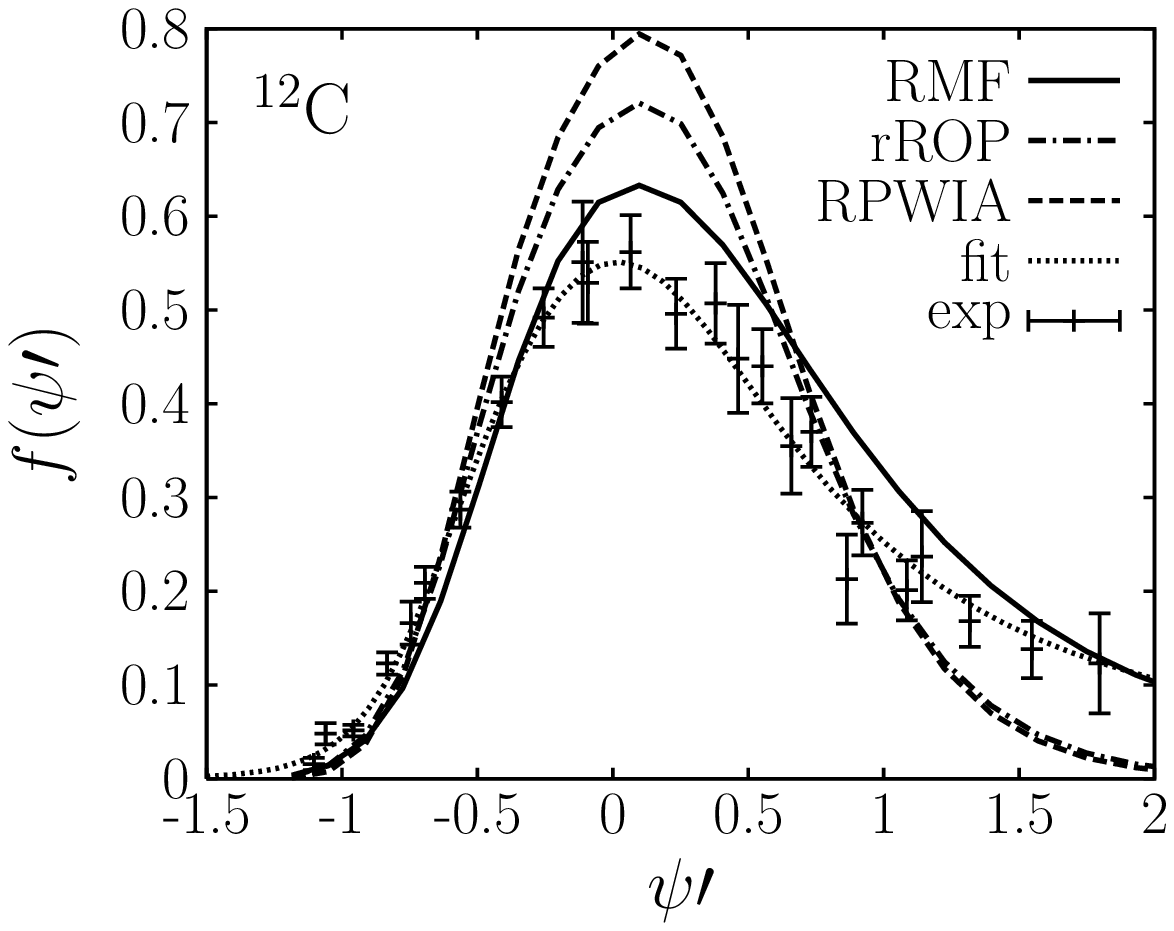}
\caption{Scaling function for $^{12}$C$(e,e')$ evaluated with the RPWIA (dashed line), rROP (dot-dashed)
and RMF (solid) approaches compared to the experimental function together with a phenomenological
parameterization (dotted).}
\label{exp_fsi_cc2}
\end{center}
\end{figure}

\begin{figure}[htb]
\begin{center}
\includegraphics[scale=0.6]{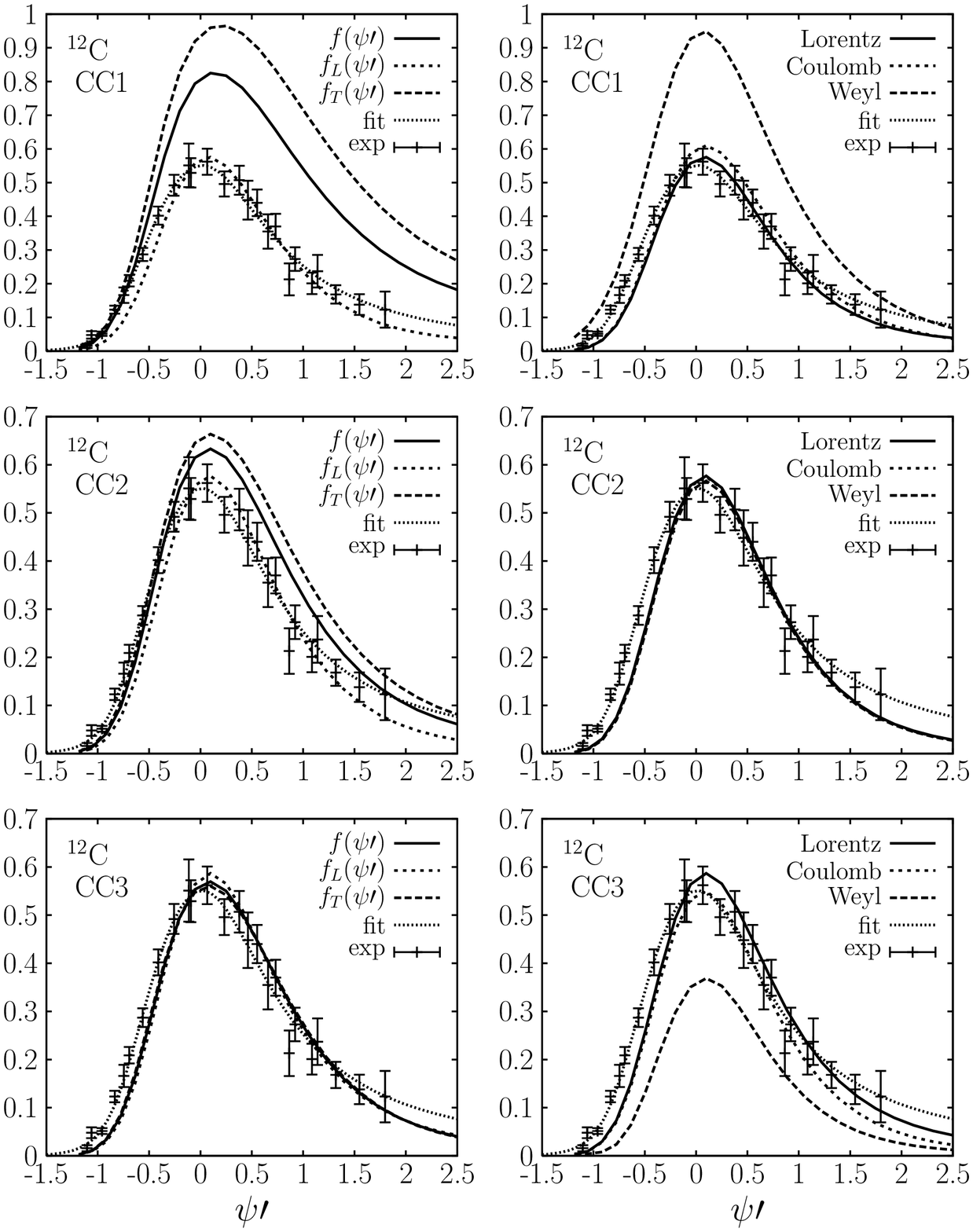}
\caption{Scaling functions evaluated for different currents and gauges compared to experiment.
Left panels show $f(\psi')$ as well as the contributions $f_L(\psi')$ and $f_T(\psi')$ for the
three choices of the current: CC1 (top), CC2 (middle) and CC3 (bottom) and the gauge fixed
to Lorentz. Right panels show the results for the longitudinal scaling function $f_L(\psi')$
for the three currents and the three gauges: Lorentz (solid line), Coulomb (short-dashed) and Weyl
(dashed).}
\label{f_gauge_rmf}
\end{center}
\end{figure}

A comparison between the theoretical superscaling functions and the averaged QE phenomenological
function obtained from the analysis of $(e,e')$ data is presented in Fig.~\ref{exp_fsi_cc2}. 
We have selected the CC2 operator and show results for the three different descriptions 
of the continuum final state, namely, the plane wave limit (dashed line), rROP (dot-dashed)
and RMF (solid). Lorentz gauge has been assumed.
The symmetric character of the RPWIA and rROP curves differs clearly from the 
experimental analysis. On the contrary, the RMF approach displays an asymmetric shape with a long
tail extended to positive values of the scaling variable $\psi'$ which follows closely the
behaviour of the phenomenological function. As already mentioned in previous work~\cite{PRL},
this asymmetry of the RMF description constitutes a basic difference from other models presented
in the literature~\cite{Antonov1,Antonov2,Antonov3} where the long tail in $f(\psi')$ is largely absent.

The asymmetry in data has usually been ascribed to ingredients beyond the mean field, such as
short-range correlations, induced non-localities and two-body currents. 
Within a non-relativistic aproach such ingredients are needed in order to
get the asymmetry~\cite{Co,Ble01,Nie04}.
However, here we show that a large amount of the asymmetry 
is indeed obtained within the framework of the RIA and 
a RMF description of the final continuum nucleon states. 
This is in accordance with some previous works~\cite{Dirac1,Dirac2,Corr,Sharma} 
where a comparison between Dirac-Brueckner-Hartree-Fock (DBHF) calculations and
Dirac-Hartree ones indicates that
effects from correlations and Fock terms in the DBHF calculation can be 
accounted for by the simple Dirac-Hartree approach fitted to saturation properties 
of nuclear matter.
This is at variance with the non-relativistic mean field case.
In this respect, note that the Dirac equation
in presence of scalar and vector local potentials can be reduced to a 
non-relativistic Schr\"odinger-like
equation with energy-dependent and non-local terms~\cite{Udias2}.
Results in Fig.~\ref{exp_fsi_cc2} show
that the asymmetry in the scaling function
can be produced via local, energy-independent relativistic potentials within
the impulse approximation, and moreover, such asymmetry is very close to the experiment. 
This outcome does not contradict the additional role that may play correlations and exchange
currents not accounted for within the relativistic mean field calculation.
However, the small magnitude of the local central and spin-orbit potentials 
involved in the non-relativistic approach cannot yield a significant
asymmetry. Only a strong non-locality of the potentials (effective values of the 
mass and energy) may give rise
to an {\it asymmetric} differential cross section~\cite{Udias-asymmetry}.

To complete the analysis, in Fig.~\ref{f_gauge_rmf} (left panels) we select the RMF description 
of FSI and compare data and the fit curve with the
theoretical results for the three choices of the current operator: CC1 (top panel), CC2 (middle)
and CC3 (bottom). In each case we also present a separate analysis of the $L$ and $T$ channels
involved in the process. As already shown in previous results, the longitudinal
contribution $f_L(\psi')$ does not depend on the current operator choice and agrees nicely with
experiment. On the contrary, the strength of the transverse response with the CC1 current leads
to a function $f_T(\psi')$ that is roughly twice the data. Discrepancies between $f_L(\psi')$
and $f_T(\psi')$ are mild (negligible) for CC2 (CC3) leading to a global scaling function
$f(\psi')$ which is in accord in both cases with the experiment. 
All the curves presented for the longitudinal 
contribution $f_L(\psi')$ satisfy the Coulomb sum rule, i.e., they integrate to unity.

Up to present we have only discussed calculations corresponding to the Lorentz gauge. 
Results for the Coulomb gauge are similar for
all FSI descriptions whereas the Weyl gauge leads in most of the cases 
to very important discrepancies.
This is clearly illustrated in the right panels of Fig.~\ref{f_gauge_rmf} where we show
the longitudinal scaling function $f_L(\psi')$ evaluated for
the three gauges: Lorentz (solid line), Coulomb (short-dashed) and Weyl (dashed),
and the three current operator choices: CC1 (top panel), CC2 (middle) and CC3 (bottom). The
RMF approach for the final state has been assumed. The discussion of results follows
in general similar trends for the rROP model.
Note that the gauge does only affect the longitudinal response. As observed in Fig.~\ref{f_gauge_rmf},
Lorentz and Coulomb gauges lead to very close results for all currents, particularly for CC2, where
there is no distinction between the two curves. In the case of the Weyl gauge, an important
difference emerges between the CC2 operator and the other two, CC1 and CC3.
In the former, the scaling function $f_L(\psi')$ 
is almost the same for the three gauges. This is in accordance with the fact that the
continuity equation is highly fulfilled within the RMF model and the CC2 current operator. 
By contrast, the CC1 and CC3 operators within the Weyl gauge give rise to longitudinal
scaling functions which depart significantly from data being much larger (smaller) 
for CC1 (CC3). In both cases the Coulomb sum rule is clearly violated. 
These results reinforce our confidence in the
adequacy of descriptions of inclusive $(e,e')$ reactions when based on the 
RMF-FSI approach and the CC2 current operator. 
Only in this case the results do not depend on the specific gauge 
selected\footnote{Although not shown, the Weyl
gauge leads to different results when the rROP model is assumed 
independently of the current operator selected.}
and moreover, they are also shown not to be modified by the dynamical enhancement 
of the lower components~\cite{Udias-asymmetry}.

\subsection{Inclusive QE charged-current $(\nu,\mu)$ reactions}

\begin{figure}[htb]
\begin{center}
\includegraphics[scale=0.8]{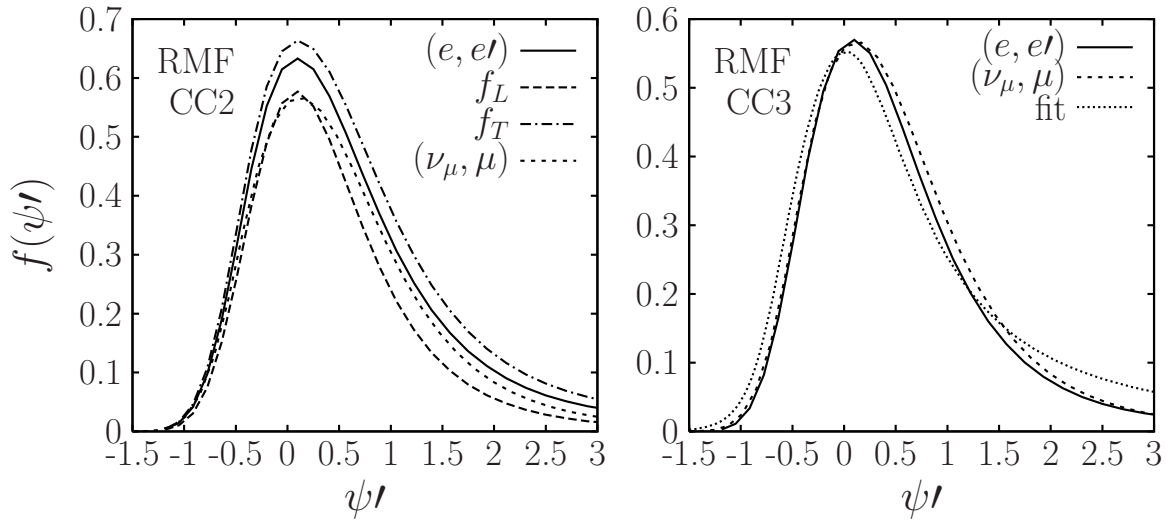}
\caption{Consistency of scaling functions evaluated for $(e,e')$ and $(\nu,\mu)$ reactions.
The incident electron (neutrino) energy is fixed to 1 GeV and the scattering angle to $45^0$.
Left panel refers to results obtained with the CC2 current (electron scattering) and a
separate $L$ and $T$ contribution is presented. Right panel shows results for CC3 choice.}
\label{electron_neutrino}
\end{center}
\end{figure}

The analysis of scaling and superscaling for CC neutrino-nucleus reactions has been presented
in ~\cite{neutrino1,PRL}. It is important to point out that any reliable calculation on
neutrino-nucleus cross sections requires first to be tested against electron scattering data. 
Hence two different approaches can be pursued.
First, using the scaling behaviour of $(e,e')$ cross sections and
the universality property of the superscaling function, we can make 
predictions for inclusive $(\nu,\mu)$ 
reactions by taking the empirical electron scattering scaling function $f_{exp}(\psi')$. 
This strategy, applied not only to the QE regime but also to $\Delta$ kinematical region, was
analyzed at depth in~\cite{neutrino1}. The second approach, considered in \cite{PRL}, consists of
evaluating explicitly $f(\psi')$ for $(\nu,\mu)$ reactions within a specific model, 
namely, RDWIA. The scaling function obtained in this way can
be compared directly with the model predictions given for $(e,e')$ processes. This allows us to
check the scaling behaviour of the calculations and moreover, the consistency of the universality
assumption of $f(\psi')$ and also the capability of the model to reproduce the
experimental data.

In~\cite{PRL} we presented a detailed investigation of CC neutrino-nucleus scattering reactions
within the RIA framework. We proved that superscaling is verified to high accuracy 
by the model calculations even
in presence of strong relativistic potentials. Importantly, the results obtained when FSI were described
by means of the RMF potential presented the right asymmetry compared with data. This is 
fully consistent with the discussion outlined in previous section concerning the
study of inclusive $(e,e')$ reactions. This consistency is clearly illustrated in
Fig.~\ref{electron_neutrino} where we compare the scaling functions evaluated from
$(e,e')$ and $(\nu,\mu)$ reactions. We only consider the RMF-FSI case as this is the only model which
is in accordance with data. However, the consistency between electron and neutrino scattering
calculations applies also to rROP and RPWIA approaches. In the left panel of 
Fig.~\ref{electron_neutrino} the CC2 prescription has been assumed for $(e,e')$ and the separate
contribution of both channels, $L$ and $T$, is also shown. For completeness, the case of the 
CC3 operator is
considered in the right panel where the fit curve to experiment is also drawn for reference. From
these results, it is clear that the universality assumption of the scaling function is highly
fulfilled within the present RIA model. This supports the use of the experimental
$(e,e')$ scaling function in order to predict reliable 
neutrino-nucleus scattering cross sections~\cite{neutrino1}.

Although not presented for simplicity, a separate analysis of the various channels, $L$, $T$ and $T'$,
contributing to $(\nu,\mu)$ reactions shows that $f_T(\psi')=f_{T'}(\psi')=f(\psi')$, i.e.,
scaling of zeroth kind is verified. The longitudinal contribution leads to a scaling
function $f_L(\psi')$ which departs significantly from $f(\psi')$. However, one should be
cautious because the $L$ contribution to inclusive
$(\nu,\mu)$ cross sections is almost negligible compared with the transverse, $T$, $T'$, ones.


\begin{figure}[htb]
\begin{center}
\includegraphics[scale=0.7]{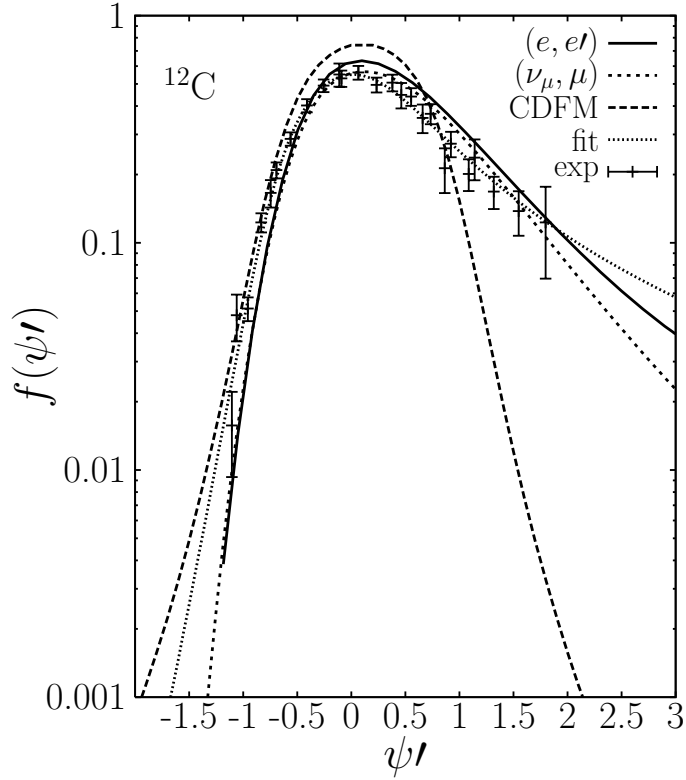}
\caption{As in previous figure but in logarithmic scale. We also include data and compare with the
predictions given by the CDFM model (see text for details).}
\label{electron_neutrino_log}
\end{center}
\end{figure}

\begin{figure}[htb]
\begin{center}
\includegraphics[scale=0.8]{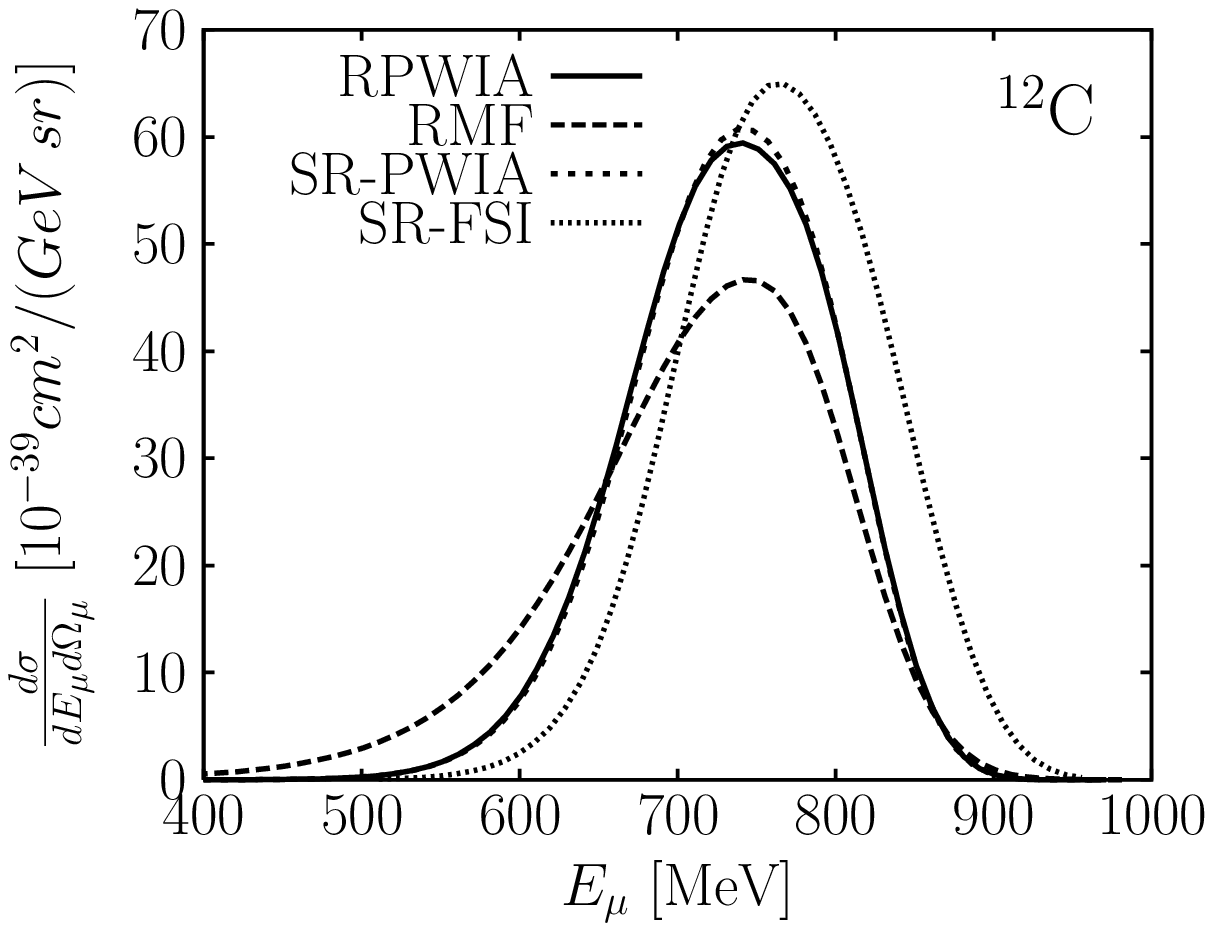}
\caption{Predictions for $^{12}$C$(\nu,\mu)$ differential cross sections. The RIA model
is compared with the SR approach. Two different descriptions of FSI have been
considered. First the plane wave limit which leads to very similar results. Second FSI described
by means of the RMF model (dashed) and making use of a Woods Saxon potential in the SR framework
(dotted).}
\label{ja_quique}
\end{center}
\end{figure}

To conclude with the analysis of results, we present again in Fig.~\ref{electron_neutrino_log}
the scaling function predictions for both $(e,e')$ and $(\nu,\mu)$ reactions with the
RMF description in the final state, and compare with data and the fit curve. Here we use a 
logarithmic scale in $f(\psi')$ in order to enlarge the discrepancies between theory and
experiment in the scaling region, i.e., negative $\psi'$-values. In this region, our model
predictions for electron and neutrino processes are in fully agreement and tend to underpredict the 
data. This is not unexpected because the model is entirely based on one-body phase space.
Ingredients beyond the impulse approximation, i.e.,
multinucleon knockout, either induced by exchange currents, correlations
or rescattering effects, are surely needed to get more strength in the scaling function 
which will be then closer to the experiment. 
This is in fact the case of the Coherent Density Fluctuation Model 
(CDFM) for correlations presented in~\cite{Antonov1,Antonov2,Antonov3}. This model is an extension of the RFG 
applied to finite nuclei and its prediction is also shown for comparison in 
Fig.~\ref{electron_neutrino_log}. The CDFM result, compared with the RMF calculations,
presents more strength in the negative $\psi'$-region being closer to data. However, 
CDFM is manifestly symmetrical around the QE peak.\footnote{The author is aware of a new
development of the CDFM model in which asymmetry is incorporated in an effective
way~\cite{Anton}.} As discussed in previous sections, the
asymmetry of the scaling function comes mainly from the inclusion of FSI in the reaction
mechanism. This ingredient, when based on the RMF description, gives rise to the
right asymmetry shown by the experiment.

A detailed analysis of scaling and cross sections within a non-relativistic
mean field approach has been presented
in~\cite{neutrino2}. In Fig.~\ref{ja_quique} we compare CC neutrino-nucleus cross sections obtained
with the RIA model and the semirelativistic (SR) one
developed in~\cite{neutrino2}. The SR approach includes important relativistic ingredients
in the current operator. This has been already illustrated in previous 
works~\cite{Amaro:2002mj,neutrino2,Amaro96,Amaro96b}
and it is also clearly shown up by results in Fig.~\ref{ja_quique}
corresponding to the plane wave limit;
the fully relativistic and the SR calculations 
lead to very similar curves in spite of
using different descriptions of the initial bound nucleon wave functions: solutions of
the Dirac-Hartree model and the Schr\"odinger equation with a Woods Saxon
potential, respectively. On the contrary, relativistic and SR results are very different when
FSI are included by using potentials (RMF and Woods Saxon) in the final state.
Whereas the RMF shows a significant asymmetry,
the SR-FSI approach shifts the SR-PWIA result to higher muon energies, but maintains
the global symmetry of the cross section~\cite{Amaro:2001xz,neutrino2}.
Asymmetry within the non-relativistic framework requires ingredients 
in the final state beyond the impulse approximation and mean field picture:
correlations, non-localities, exchange terms, multinucleon emission.
By contrast, most of the asymmetry involved by the experiment is already
obtained within a simple, fully relativistic mean field model based on the impulse approximation and 
describing FSI by means of strong local scalar and vector potentials. This makes a crucial difference
between both, relativistic and non-relativistic, descriptions.

\section{Conclusions}

In this paper we have presented a global study of scaling and superscaling properties
concerning inclusive QE electron and CC neutrino-nucleus scattering reactions. The
general framework in which the calculations have been performed is the RIA. 
This is a simplified description of the reaction mechanisms
because it presupposes that the processes to be calculated depend entirely on one-body
phase space. However, in spite of its simplicity, the RIA has been used to
describe successfully QE $(e,e'N)$ reactions. For the inclusive $(e,e')$ and $(\nu,\mu)$
processes we are interested in, a crucial ingredient is the description of FSI. We have
considered different options consistent with retaining in the calculations 
the contribution from the inelastic channels. This is provided in our model by using
purely real potentials. Three different cases have been explored:
i) the plane wave limit, ii) using the real part of the energy-dependent relativistic optical
potentials parameterized by Clark et al.~\cite{Clark}, and iii) employing
the same relativistic mean field potential considered in the description of 
the initial bound nucleon states.

The differential $(e,e')$ cross sections are shown to be very sensitive to some
ingredients of the model, particularly, the specific current operator selected 
(CC1 vs CC2) and the relativistic FSI description. 
The former, off-shell effects, enter essentially in the
transverse channel, whereas the longitudinal one shows a very mild dependence 
with the current operator choice. 
Concerning FSI, a basic difference emerges when comparing the rROP and RMF approaches to the
cross section. The latter presents a significant asymmetry extended to higher values of the
transfer energy. The origin of this effect
is directly linked to the use of very strong relativistic
optical potentials~\cite{Udias-asymmetry}.

The main goal of this paper is centered in the analysis of the scaling behaviour of the cross
sections and separate responses. Our results show that scaling of zeroth order is verified
for the CC2 and CC3 current operators; however, the CC1 leads to an
important breakdown of scaling because of the extremely large contribution given by the
transverse response. This outcome is strictly true when FSI effects are taken into account.
In the plane wave limit (RPWIA), results present a mild dependence with
the current operator. Concerning first kind scaling, i.e., independence on the transfer momentum, 
all our results satisfy this property which is in accord with
the general behaviour of data. Being more precise, only in the case of FSI
described with the RMF potential, the superscaling function shows some dependence
with the transfer momentum. However, this scaling breakdown is compatible with $(e,e')$ data.
A separate analysis of the two channels involved in the calculations leads also to similar conclusions.

A special comment should be made concerning second kind scaling, i.e., independence of the
nuclear species. Recent analyses of $(e,e')$ data conclude that this property is excellent,
showing only a small scaling violation in the region above the QE peak (positive $\psi'$-values).
This means that any theoretical model which does not satisfy this behaviour should be dismissed,
or at least, clearly disfavoured. In our model, the analysis performed shows that the use
of the CC1 operator leads to a visible breakdown of second kind scaling. This
applies to both FSI descriptions, rROP and RMF, and it comes totally from the transverse
response. These results strongly favour the use of the CC2 (likewise CC3) operator,
which agrees with some general comments made in the past concerning the analysis of 
{\it exclusive} QE $(e,e'p)$ data at high missing momentum values~\cite{highp}.
Note that second kind scaling is highly fulfilled for the longitudinal contribution $f_L(\psi')$.

A comparison with the {\it phenomenological} scaling function extracted from $(e,e')$ data
shows that only the FSI description based on the RMF approach leads
to the right amount of asymmetry. Within the one-body context,
this significant shift of strength to positive $\psi'$-values 
only occurs in presence of strong scalar and vector relativistic optical
potentials. This is the case of the energy-independent RMF approach. On the contrary,
the energy-dependent scalar and vector terms in the rROP are
importantly reduced for increasing outgoing nucleon energies (positive $\psi'$-region).
Comparison with data indicates that the RMF-FSI model
incorporates important dynamical effects missed by other models. 
In a non-relativistic approach,
the asymmetry can be only generated by including ingredients beyond the 
impulse approximation. Here we show that a simple local, energy-independent relativistic
mean field potential takes care of the basic behaviour presented by $(e,e')$ data.

The gauge analysis performed in this work leads to an important conclusion concerning
the RMF-FSI approach. Notice that gauge ambiguity only affects the longitudinal contribution 
$f_L(\psi')$. Our study shows that Coulomb and Lorentz gauges give rise to very similar results, 
being in general significantly different
from the Weyl ones. However, in the particular case of FSI
described with the RMF potential and the CC2 current operator selected,
the longitudinal scaling function does not show sensitivity to the gauge election. Only in
this situation the results of $f_L$ are very similar for the three gauges and, more importantly,
all of them essentially satisfy the Coulomb sum rule, i.e., they integrate to unity. 
By contrast, for CC1 and CC3 prescriptions,
the Weyl $f_L$ results deviate significantly from the other gauges breaking also the
Coulomb sum rule. This general result gives us confidence in the adequacy of
descriptions of QE $(e,e')$ reactions when based on the RMF-FSI approach and the CC2 current operator. 
This choice, in addition to be mostly insensitive to gauge ambiguities and dynamical relativistic
effects, leads to the {\it correct} asymmetry shown by data.

To conclude, we have compared the scaling function evaluated from $(e,e')$ calculations with 
that obtained from $(\nu,\mu)$ reactions. The results for both processes are very close,
and this is fully consistent with the
general statement on the universality property of the scaling function.
Within the RIA framework, the superscaling function is basically the same
for the two $t$-channel processes considered. This supports
the general approach considered in~\cite{neutrino1}, i.e., the 
use of the scaling function extracted from $(e,e')$
data to predict neutrino-nucleus cross sections. It will be very interesting in the
future to analyze if the universality property emerges also from 
RIA calculations applied to NC neutrino-nucleus scattering reactions~\cite{NC}.

\section*{Acknowledgements}
This work was partially supported by funds provided by Ministerio de Educaci\'on
y Ciencia (Spain) and FEDER funds, under Contracts Nos.
FPA2005-04460 and FIS2005-01105, by the
Junta de Andaluc\'{\i}a, and by the INFN-CICYT collaboration
agreements N$^0$ 05-22.
This work has benefited from very fruitful
discussions with J.E. Amaro, M.B. Barbaro, 
T.W. Donnelly, E. Moya de Guerra and J.M. Ud\'{\i}as. I warmly thank all of them.
I also acknowledge A.N. Antonov for providing the CDFM scaling function.


\end{document}